\documentclass[preprint,showpacs,preprintnumbers,amsmath,amssymb, superscriptaddress,longbibliography,nofootinbib]{revtex4-1}
\usepackage{graphicx} 
\usepackage{xcolor}
\usepackage{amsmath}
 \usepackage{hyperref}
\usepackage[top = 2cm, bottom = 2cm, right = 2cm, left =2cm]{geometry}

\usepackage{tikz}
\usetikzlibrary{decorations.pathmorphing,calc}
\begin{document}

\title{An Analytical Two-Incompressible-Fluid Star with a Mixed Ordinary–Dark Matter Core and an Ordinary-Matter Envelope}

\author{Milko Estrada}
\email{milko.estrada@gmail.com}
\affiliation{Departamento de Física, Facultad de Ciencias, Universidad de Tarapacá, Casilla 7-D, Arica, Chile}

\author{Santiago Esteban Perez Bergliaffa}
\email{sepbergliaffa@gmail.com}
\affiliation{Departamento de Fíısica Teórica, Instituto de Física, Universidade do Estado de Rio de Janeiro, CEP 20550-013, Rio de Janeiro, Brazil}

\date{\today}

\begin{abstract}
We construct an analytical relativistic two-fluid star characterized by
a mixed core, where ordinary matter and dark matter coexist as two
independently conserved incompressible perfect fluids, and an envelope
composed exclusively of ordinary matter. The fluids exchange neither
matter nor energy and interact only through the common spacetime geometry,
with the ordinary component extending across the core--envelope interface
while the dark component is confined to the core. Despite the
mixed-core--single-fluid-envelope structure and the internal interface,
the system remains analytically tractable, allowing us to obtain explicit
expressions for the pressures and metric functions and to follow directly
the effects of the dark-matter fraction and relative core size. We
determine the physically admissible parameter space and derive a
Buchdahl-like critical compactness associated with the divergence of the
central pressure, whose value depends on the relative dark-matter density
and the size of the mixed core. The Schwarzschild constant-density star
and its standard critical value, $2M/R=8/9$, are recovered in the
corresponding one-fluid limit. The mass--radius analysis further shows
that configurations with the same global compactness can correspond to
distinct internal matter distributions. Beyond providing an analytically
controlled description of a core-confined second component, the
construction offers a useful benchmark for identifying qualitative trends
that may subsequently be examined in more realistic
dark-matter-admixed neutron-star models, whose detailed treatment lies
beyond the scope of the present work.

\end{abstract}

\maketitle

\section{Introduction}

Since its discovery in 1916, the so-called Schwarzschild interior solution (SIS) \cite{Schwarzschild1916}, describing a static and isotropic perfect-fluid configuration with constant density, has provided one of the simplest analytical models for investigating the relativistic structure of compact stars. Despite the idealized character of the constant-density assumption, the SIS remains useful both as a toy model and as a reference configuration against which more involved stellar models can be compared. In fact, the interiors of sufficiently dense stars, such as neutron stars, may in certain regimes be approximated as having an approximately uniform density, lending practical interest to the SIS \cite{Lemos:2014lza}. Moreover, as discussed in \cite{Arbanil:2014usa}, a constant-density description can be associated with matter that is essentially incompressible and capable of sustaining extremely high pressures. Such conditions may arise when the characteristic particle velocities become relativistic, namely when the temperature becomes comparable to the particle rest mass for bosonic matter or when the Fermi energy approaches the rest mass for fermionic matter. In these regimes, the density approaches approximately one particle per cubic Compton wavelength \cite{Arbanil:2014usa}. The incompressible model also leads to the well-known Schwarzschild compactness limit, which provides a useful benchmark for comparison with other compactness bounds. In this context, electrically charged incompressible stellar configurations, extending the standard SIS, were investigated in \cite{Arbanil:2014usa}.

A well-known subtlety of the constant-density description concerns the speed of sound. A strictly incompressible equation of state formally corresponds to an infinite speed of sound. Nevertheless, it has been argued that this feature need not substantially affect the global stellar structure described by the constant-density model \cite{Arbanil:2014usa}. A complementary point of view was recently discussed in \cite{deAguiarAlves:2024yer}. Although the background energy density of the SIS is constant, the quantity $\left(\partial P/\partial\rho\right)$ entering a perturbative analysis may be associated with perturbations whose energy density is not constant. Within this interpretation, Chandrasekhar's pulsation equation can suggest stable Schwarzschild-star configurations in some situations. However, the authors of \cite{deAguiarAlves:2024yer} themselves emphasize that the divergence of $\left(\partial P/\partial\rho\right)$ makes such a stability interpretation delicate and that conclusions based on this procedure should therefore be regarded with caution. More generally, the requirement that perturbations propagate causally provides an important constraint on relativistic stellar matter and leads to the familiar causality bound on stellar compactness \cite{SchaffnerBielich2020}. Thus, throughout this work, the constant-density description should be understood as an analytically tractable idealization rather than as a complete microscopic equation of state for realistic stellar matter. Its purpose here is to provide a controlled setting in which the gravitational effects of a second, independently conserved matter component can be isolated and studied analytically.

The analytical simplicity of the SIS has made it a useful setting for investigating a broad range of relativistic phenomena. Its stability has been discussed in \cite{Chandrasekhar1964,Konoplya2019}, while the inclusion of slow rotation was considered in \cite{chandra1974}. The gravastar limit of the SIS was investigated in detail in \cite{Mazur2015,Posada2018}\footnote{For the generalization to the slowly rotating case, see \cite{Posada2016}.}. Constant-density relativistic fluid spheres supported by thin shells were constructed in \cite{Rosa:2020hex}. In particular, configurations were obtained either by collapsing the external layers of the Schwarzschild constant-density star into a thin shell and matching to an exterior Schwarzschild spacetime, or by introducing an internal vacuum region through matching with a Minkowski geometry. These models can attain compactness values arbitrarily close to the black-hole limit while remaining regular at the center, providing examples that can exceed the usual Buchdahl bound \cite{Rosa:2020hex}. The SIS has also been employed to investigate approximate universal relations between neutron-star (NS) properties, i.e., relations that are largely independent of the equation of state (EOS) \cite{Saes2024}, to discuss the positivity of the mass \cite{deAguiarAlves:2024yer}, and to calculate the self-force in non-vacuum regions \cite{Seenivasan2025}. More recently, axial gravitational perturbations of uniform-density stars in an Anti-de Sitter (AdS) spacetime background were studied in \cite{Lin2025}. These examples illustrate how the SIS continues to serve as a simple analytical laboratory in which additional physical ingredients can be incorporated and their effects isolated.

The peculiar properties of the Schwarzschild star have also motivated investigations of possible observational signatures. Recent developments in very long baseline interferometry, particularly those associated with the Event Horizon Telescope (EHT), have demonstrated the possibility of probing the strong-field region of compact objects. In this context, the interferometric pattern associated with relativistic images around a Schwarzschild star was investigated in \cite{Gao:2024ksc}. In contrast with a Schwarzschild black hole, the Schwarzschild star considered there possesses both inner and outer relativistic images associated with the region inside the photon sphere, leading to a more involved interferometric pattern that could, in principle, help distinguish the two geometries \cite{Gao:2024ksc}. Although the present work does not pursue observational predictions and treats the constant-density configuration primarily as a toy model, these studies provide another example of the range of questions for which the SIS has been employed.

Of particular interest for the present work is the application of constant-density configurations to compact objects containing more than one fluid. In \cite{Zollner2022}, the SIS was employed to model the core of a compact object in a regime where uncertainties in the EOS are relevant. Two-fluid compact objects have also attracted attention in the context of dark matter (DM). In \cite{Cassing2022}, compact objects composed of two different DM fluids were systematically investigated, including both core-shell configurations, in which the fluids occupy separate regions, and mixed configurations, in which the two DM components coexist. The matter models considered there include incompressible fluids, free and interacting Fermi gases, and self-bound DM equations of state. In particular, \cite{Cassing2022} analyzed stars consisting of two layers of incompressible fluids with different constant densities. In these configurations, one fluid forms the core and the other the shell. The pressure is continuous across their interface, whereas the density can undergo a discontinuous change, corresponding to a first-order phase transition. The resulting mass-radius relations display properties absent in the corresponding one-fluid configurations, and, for suitable choices of the density ratio and core radius, highly compact objects approaching the causality limit can be obtained \cite{Cassing2022}. This illustrates that even simple constant-density models can acquire a substantially richer structure once more than one matter component is introduced.

There is also a physical motivation for considering the coexistence of ordinary matter and DM within the same compact object. In particular, asymmetric DM may accumulate inside neutron stars and contribute to their internal structure \cite{Giangrandi}. Two-fluid models in which the dark and baryonic components are coupled through gravity have been used to describe such configurations, including cases in which the accumulated DM is concentrated predominantly in the stellar core \cite{Giangrandi,Ellis2018}. Depending on the properties, abundance, and spatial distribution of the dark component, two-fluid neutron-star models may develop either DM-core- or DM-halo-dominated configurations \cite{Kumar2025}. Scenarios involving centrally concentrated DM have also been considered in \cite{Gresham2018,Ellis2018}, whereas configurations in which the dark component extends beyond the ordinary-matter radius have been discussed in \cite{Shawqi2024,Kumar2025}. These studies provide a physical motivation for considering gravitationally coupled matter components with different spatial extents, although the detailed distribution of DM depends on its microscopic properties and equation of state.

Motivated by this general picture, in this work we introduce an analytically tractable two-fluid extension of the Schwarzschild constant-density star. We consider two independently conserved incompressible fluids, which may be interpreted as ordinary matter and DM, interacting through the common spacetime geometry. The two components coexist in the central region of the object, forming a mixed core, while only the ordinary-matter component extends into the envelope. The aim of this construction is not to provide a microscopic model of a realistic dark-matter-admixed neutron star, but rather to isolate analytically the gravitational effects associated with a centrally concentrated second component and with the transition from a mixed two-fluid core to a single-fluid envelope. In this respect, the construction differs from the two-layer incompressible model considered in \cite{Cassing2022}, in which one constant-density fluid occupies the core and a different constant-density fluid occupies the shell. Here, instead, one of two initially coexisting components terminates at a finite internal radius while the other continues toward the stellar surface. This structure allows the effects of the additional component on the pressure profiles, matching conditions, global compactness, and the corresponding Buchdahl-like critical configuration to be studied within a largely analytical framework.

A further motivation for the present construction is to explore how far an analytically tractable two-fluid model can provide insight into the properties of compact stars containing an additional matter component. Although establishing a direct connection with realistic dark-matter-admixed neutron stars requires a more detailed treatment beyond the scope of this work, the availability of analytical solutions allows the role of the second component to be isolated and its effects on the internal structure and compactness to be studied explicitly. The present model may therefore provide qualitative guidance for the analysis of more realistic dark-matter-admixed neutron-star configurations.

An important feature of the construction is that, despite the presence of two matter components and two distinct stellar regions, the problem remains largely analytical. As will be shown below, the field equations can be solved explicitly to a large extent, and the standard interior Schwarzschild solution is recovered through a definite limiting procedure. This allows us to investigate explicitly how the presence of the second fluid affects the internal pressure distribution, the matching between the core and the envelope, and the compactness of the resulting configurations. In this way, the model provides an analytically controlled extension of the SIS that combines a mixed two-fluid core with a single-fluid envelope while retaining a direct connection with the standard Schwarzschild constant-density star.

\section{Structure of the two-component configuration}

Let us present a description of the model, which consists of two regions that may be referred to as the core and the envelope. 
\begin{center}
\tikzset{every picture/.style={line width=0.75pt}} 

\begin{tikzpicture}[x=0.75pt,y=0.75pt,yscale=-1,xscale=1]

\draw   (295.5,109.75) .. controls (295.5,50.79) and (343.29,3) .. (402.25,3) .. controls (461.21,3) and (509,50.79) .. (509,109.75) .. controls (509,168.71) and (461.21,216.5) .. (402.25,216.5) .. controls (343.29,216.5) and (295.5,168.71) .. (295.5,109.75) -- cycle ;
\draw   (350.5,111.5) .. controls (350.5,80.57) and (375.57,55.5) .. (406.5,55.5) .. controls (437.43,55.5) and (462.5,80.57) .. (462.5,111.5) .. controls (462.5,142.43) and (437.43,167.5) .. (406.5,167.5) .. controls (375.57,167.5) and (350.5,142.43) .. (350.5,111.5) -- cycle ;
\draw    (406.5,111.5) -- (481.51,178.17) ;
\draw [shift={(483,179.5)}, rotate = 221.63] [color={rgb, 255:red, 0; green, 0; blue, 0 }  ][line width=0.75]    (10.93,-3.29) .. controls (6.95,-1.4) and (3.31,-0.3) .. (0,0) .. controls (3.31,0.3) and (6.95,1.4) .. (10.93,3.29)   ;
\draw    (406.5,111.5) -- (415.91,103.59) -- (447.45,77.77) ;
\draw [shift={(449,76.5)}, rotate = 140.7] [color={rgb, 255:red, 0; green, 0; blue, 0 }  ][line width=0.75]    (10.93,-3.29) .. controls (6.95,-1.4) and (3.31,-0.3) .. (0,0) .. controls (3.31,0.3) and (6.95,1.4) .. (10.93,3.29)   ;

\draw (413,73) node [anchor=north west][inner sep=0.75pt]   [align=left] {$\displaystyle R_{i}$};
\draw (486,177) node   [align=left] {\begin{minipage}[lt]{68pt}\setlength\topsep{0pt}
$\displaystyle R_{e}$
\end{minipage}};
\draw (533,42) node   [align=left] {\begin{minipage}[lt]{68pt}\setlength\topsep{0pt}
Vacuum
\end{minipage}};

\end{tikzpicture}
\end{center}

The core, or Region I, is defined by $0 \le r \le R_i$. In this region, two matter components are present, characterized by constant energy densities denoted by $\rho_o$ and $\rho_D$, with corresponding pressures $p_o^{(I)}$ and $p_D$. As will be discussed below, the subscripts $o$ and $D$ refer to ordinary matter and dark matter, respectively. Within Region I, we define the total energy density and total pressure as
$\rho_T=\rho_D+\rho_o$ and $p_T^{(I)}(r)\equiv p_T(r)=p_D(r)+p_o^{(I)}(r)$.

Both regions are described by a metric adapted to a static and spherically symmetric spacetime. In particular, the metric in Region I is given by
\begin{equation} \label{MetricaRegion1}
ds^2=-\exp\left(2\Phi^{(I)}(r)\right)dt^2
+\exp\left(2\Psi^{(I)}(r)\right)dr^2+r^2d\Omega^2.
\end{equation}
In Appendix \ref{ApendiceEstrellaSch}, we present the form of the equations of motion for incompressible fluids. Under the assumptions mentioned above, and using $e^{2\Psi^{(I)}}=(1-2m_T(r)/r)^{-1}$, the relevant equations for the core are

\begin{equation} \label{ConservacionD}
\frac{d}{dr}p_D(r)
=-\left[p_D(r)+\rho_D\right]\frac{d\Phi^{(I)}(r)}{dr},
\end{equation}

\begin{equation} \label{ConservacionO}
\frac{d}{dr}p_o^{(I)}(r)
=-\left[p_o^{(I)}(r)+\rho_o\right]\frac{d\Phi^{(I)}(r)}{dr},
\end{equation}

\begin{equation}
m_o(r)=\frac{4}{3}\pi r^3\rho_o,
\label{mo}
\end{equation}
\begin{equation}
m_D(r)=\frac{4}{3}\pi r^3\rho_D,
\label{md}
\end{equation}

\begin{equation}
\label{dphi}
\frac{d\Phi^{(I)}(r)}{dr}
=\frac{m_T(r)+4\pi r^3p_T(r)}
{r[r-2m_T(r)]},
\end{equation}
with $m_T=m_o+m_D$, namely,
\begin{equation} \label{MasaTotalRegionI}
m_T(r)=\frac{4\pi}{3}\rho_T r^3.
\end{equation}

These equations must be supplemented with the appropriate boundary conditions at $r=R_i$ and $r=R_e$, as discussed below.

Equations \eqref{ConservacionD} and \eqref{ConservacionO} follow from
$\nabla_\nu \big(T^{\mu \nu}\big)_D = 0$ and $\nabla_\nu \big(T^{\mu \nu}\big)_o = 0$. We therefore consider the two fluids to be independently conserved. In particular, the two components do not exchange matter or energy directly, although both contribute to and evolve in the same spacetime geometry and hence interact gravitationally. Motivated by scenarios in which dark matter can be modeled as an additional matter component with negligible non-gravitational coupling to ordinary matter, we refer to fluid $D$ as the dark matter fluid and to fluid $o$ as the ordinary matter fluid.

In this work, we assume that the dark matter component is confined to the core and terminates at $r=R_i$. Accordingly, its pressure is required to vanish at the core boundary, $p_D(R_i)=0$, and therefore $p_T^{(I)}(R_i)=p_o^{(I)}(R_i)$. The abrupt termination of the constant density $\rho_D$ at $R_i$ should be understood as part of the incompressible-fluid idealization adopted here, rather than as a microscopic description of the transition between dark and ordinary matter. In particular, it provides a simple analytical representation of a centrally concentrated second component with a finite spatial extent. In the envelope, or Region II, defined by $R_i\leq r\leq R_e$, only ordinary matter is present, with constant energy density $\rho_o$ and pressure $p_o^{(II)}(r)$. The physical motivation for considering a centrally concentrated dark component was discussed in the Introduction.

In Region II, the system obeys the standard Tolman--Oppenheimer--Volkoff (TOV) equation, namely,

\begin{equation} \label{TOVII}
\frac{d}{dr}p_o^{(II)}(r)
=-\left[p_o^{(II)}(r)+\rho_o\right]
\frac{m^{(II)}(r)+4\pi r^3p_o^{(II)}(r)}
{r\left[r-2m^{(II)}(r)\right]},
\end{equation}

with the function $m^{(II)}(r)$ given by
\begin{equation}
m^{(II)}(r)
=\frac{4\pi}{3}\rho_o r^3
+\frac{4\pi}{3}\rho_D R_i^3 .
\label{MasaRegionII}
\end{equation}

In terms of the quantities defined above, the pressure at $r=R_i$ must satisfy
\begin{equation} \label{MatchingPresionIIyIII}
p_o^{(I)}(R_i)=p_o^{(II)}(R_i).
\end{equation}

Finally, the exterior of the star, Region III, where $r>R_e$, is described by the Schwarzschild vacuum solution. The pressure in the envelope must satisfy
\begin{equation}
p_o^{(II)}(R_e)=0.
\end{equation}
The total mass of the configuration is then given by
\begin{equation}
M=\frac{4\pi}{3}\left(\rho_D R_i^3+\rho_o R_e^3\right).
\label{MasaTotal}
\end{equation}

In the following subsections, we obtain the explicit form of the configuration in each region by solving the equations presented above subject to the corresponding boundary conditions.

In this way, we consider a model in which ordinary matter and dark matter coexist in the core (Region I), while only ordinary matter extends throughout the envelope (Region II). The assumption that the dark matter component is concentrated in the central region is motivated by scenarios in which dark matter accumulates inside compact stars. Within our effective description, the dark matter component is represented by an isotropic fluid that is independently conserved from ordinary matter but interacts with it gravitationally through the common spacetime geometry. The ordinary-matter component, in turn, extends from the center to the stellar surface. The resulting configuration therefore describes a compact object with a mixed ordinary-matter--dark-matter core and a purely ordinary-matter envelope. This structure provides a simple two-region realization of centrally concentrated dark matter and, at the same time, differs from configurations in which the dark matter component extends beyond the ordinary-matter radius and forms a halo.

On the other hand, as pointed out in \cite{Kumar:2021hiz}, considering static perfect-fluid spheres provides an effective approximation for modeling compact stars consistent with observational data, with the perfect-fluid description requiring isotropic pressure, \textit{i.e.}, equality between the radial and tangential pressures. In this context, as noted in \cite{Nashed:2023fpf}, the introduction of anisotropic pressures is not always physically justified. For instance, within a compact star, although the radial pressure vanishes at the stellar surface, the tangential pressure may remain nonzero in an anisotropic configuration. While such a nonvanishing tangential pressure does not by itself break spherical symmetry, it may be associated with internal fluid flows and possible dynamical imbalances or instabilities \cite{Riazi:2015hga}. Therefore, in the present work we restrict our attention to isotropic perfect fluids, which also allows us to retain the analytical simplicity characteristic of the Schwarzschild constant-density model.

It is also worth mentioning that a related scenario has recently been
considered in Ref.~\cite{Zollner:2026pxc}, where constant-density
two-fluid Schwarzschild solutions were studied, including configurations
with a mixed core surrounded by a single-fluid envelope. Despite this
structural similarity, the physical specification of the complete
configuration, its mathematical implementation, and the properties
investigated here are different. In Ref.~\cite{Zollner:2026pxc}, the
densities and central pressures of the two fluids are taken as input
parameters, and their respective radii are determined by the zeros of
their pressure profiles. In the present construction, instead, the
second component is explicitly confined to the core by imposing
$p_D(R_i)=0$, while the ordinary-matter component extends continuously
across the interface and vanishes only at the stellar surface,
$p_o(R_e)=0$. The core and envelope are consequently treated as a
single matched configuration, for which the integration constants and,
crucially, the central total pressure $p_{TC}$ are determined by the
boundary and matching conditions rather than introduced as independent
parameters. This closure naturally leads to a description in terms of
$f=\rho_D/\rho_o$, $\alpha=R_e/R_i$, and $C_i$, allowing us to map the
statically admissible parameter region and determine a Buchdahl-like
critical boundary $C_i^{\rm crit}(f,\alpha)$ associated with the
divergence of the central pressure, together with an analytical
approximation to its behavior. We further relate this critical structure
to the metric properties and mass--radius sequences, showing, in
particular, that different internal compositions and relative core sizes
can correspond to the same global compactness.

\subsection{Region I: The Core, $0 \le r \le R_i$}

In order to find the expressions for $p_D(r)$ and $p_o^{(I)}(r)$, we substitute Eq.~\eqref{dphi} into Eqs.~\eqref{ConservacionD} and \eqref{ConservacionO}, thus obtaining

\begin{equation} \label{TOVD}
   \frac{d}{dr} p_D(r)  = -\left[p_D(r) + \rho_D\right] \frac{m_T(r) + 4\pi r^3 p_T(r)}{r \left[ r - 2m_T(r) \right]}.
\end{equation}

\begin{equation} \label{TOVo}
   \frac{d}{dr} p_o^{(I)}(r)  = -\left[p_o^{(I)}(r) + \rho_o\right] \frac{m_T(r) + 4\pi r^3 p_T(r)}{r \left[ r - 2m_T(r) \right]}.
\end{equation}
By adding these equations, it follows that 
\begin{equation} \label{TOVT}
   \frac{d}{dr} p_T(r)  = -\left[p_T(r) + \rho_T\right] \frac{m_T(r) + 4\pi r^3 p_T(r)}{r \left[ r - 2m_T(r) \right]}, 
\end{equation}
which has the form of the usual TOV equation (see Eq.\eqref{TOVtotal}). Its solution is given by
\begin{equation} \label{presionTotalI}
p_T(r)=\rho_T \:\frac{-1+\kappa_T\sqrt{
\beta(r)
}}{3-\kappa_T\sqrt{\beta(r)}},
\end{equation}
where $\beta(r)\equiv 1-\frac{8}{3}\pi r^2 \rho_T$, and the dimensionless integration constant $\kappa_T$ can be written, using Eq. \eqref{presionTotalI},  as follows:
\begin{equation} \label{kappaTregionI}
    \kappa_T= \frac{3p_{TC}+\rho_T}{p_{TC}+\rho_T},
\end{equation}
where $p_{TC}\equiv p_T(0) $.
Hence the total pressure in the core (Eq.\eqref{presionTotalI}) is given by :
\begin{equation} \label{PresionTotalI1}
     p_T(p_{TC},\rho_T,r)= \rho_T \cdot   \frac{
    -(p_{TC}+\rho_T)+(3p_{TC}+\rho_T) \sqrt{\beta(r)}
}{
   3(p_{TC}+\rho_T)-(3p_{TC}+\rho_T) \sqrt{\beta(r)
}},
\end{equation}
and satisfies $p_T(0) = p_{TC}$. The expression inside the parentheses on the left-hand side of the last equation indicates the dependence on the parameter space $(p_{TC}, \rho_T = \rho_o + \rho_D)$ and on the radial coordinate $r$. With $p_T(r)$ given by Eq.~\eqref{PresionTotalI1}, Eq.~\eqref{TOVD} can be integrated to yield the corresponding expression for the dark matter pressure in the core
\begin{equation}
\label{pd}
p_D(\kappa_D,p_{TC},\rho_o, \rho_D,r)=   \kappa_D \sqrt{\rho_T} \cdot \frac{\sqrt{3+\kappa_T^2 \:\beta(r)+ 2 \sqrt{3} \kappa_T \sqrt{\beta(r)}} }{\sqrt{2 \pi }  \left ( 3- \kappa_T^2\:  \beta(r)\right )}
- \rho_D
\end{equation}
where the expression inside the parentheses on the left-hand side of the last equation indicates the dependence of the dark matter pressure on the parameter space $(\kappa_D, p_{TC}, \rho_o, \rho_D)$ and on the radial coordinate $r$. Here, as mentioned, $\rho_T = \rho_o + \rho_D$, 
and the integration constant $\kappa_T$, which has dimensions of $L^{-1}$ in geometric units (where $L$ represents length)
is given by Eq.~\eqref{kappaTregionI}.

As discussed earlier, since dark matter is present only in the core, we impose that $p_D(r = R_i) = 0$. It follows from Eq.~\eqref{pd} that
\begin{align}
\kappa_D= \sqrt{\frac{2\pi}{\rho_T}}\: \rho_D \cdot \frac{ 3-  \kappa_T^2\:\beta_i}{\sqrt{3+\kappa_T^2 \beta_i+ 2 \sqrt{3} \:\kappa_T \sqrt{\beta_i}}},
\end{align}
where $\beta_i\equiv\beta(r=R_i)$.
Hence, the dark matter pressure can be written as
\begin{align} \label{PresionDark}
    &p_D(R_i,p_{TC},\rho_o, \rho_D,r)= \rho_D \cdot 
    \left[
    \sqrt{ \frac{3+\kappa_T^2 \beta(r)+ 2 \sqrt{3} \kappa_T \sqrt{\beta(r)}}{3+\kappa_T^2\: \beta_i+ 2 \sqrt{3} \kappa_T \sqrt{\beta_i}}} \cdot \frac{3- \kappa_T^2\:\beta_i}{3- \kappa_T^2\:\beta(r)}- 1 \right],
\end{align}
with $\kappa_T$ 
given in Eq.\eqref{kappaTregionI}. It is straightforward to verify that at $r = R_i$ the dark matter pressure vanishes. 

As we will see below, the central pressure $p_{TC}$ is not a free parameter; rather, its value is determined by the boundary conditions in regions I and II. We will further show that applying these conditions allows us to test the behavior of the dark matter pressure by examining the ratios between the ordinary and dark matter densities, $\rho_o$ and $\rho_D$, and between the inner and outer radii, $R_i$ and $R_e$, respectively.

Finally, the metric coefficients in region I take the form 
\begin{equation}
\label{phipc1}
\exp(2\Phi^{(I)})= \kappa_I
\left(
1-
\frac{3p_{TC}+\rho_T}{3(p_{TC}+\rho_T)}
\sqrt{\beta(r)}
\right)^2 .
\end{equation}
where $\kappa_I$ is an integration constant.

\subsection{Region II: The Envelope, $R_i \le r \le R_e$}

As mentioned earlier, only ordinary matter is present in this region. In order to solve Einstein's equations in Region II, we consider the following line element:
\begin{equation}
\label{metric}
ds^2=-\exp\left[2\Phi^{(II)}(r)\right]dt^2
+\exp\left[2\Psi^{(II)}(r)\right]dr^2+r^2d\Omega^2 .
\end{equation}
With the mass function given by Eq.~\eqref{MasaRegionII}, we have
\begin{equation}
\exp\left[-2\Psi^{(II)}(r)\right]
=1-\frac{8\pi}{3}\rho_o r^2-\frac{8\pi}{3}\frac{\bar C}{r}.
\end{equation}
Imposing the continuity of $g_{rr}$ at $r=R_i$, it follows that $\bar C=\rho_D R_i^3$, and
\begin{equation}
\label{grr}
\exp\left[-2\Psi^{(II)}(r)\right]
=1-\frac{8\pi}{3}\rho_o r^2
-\frac{8\pi}{3}\frac{\rho_D R_i^3}{r}.
\end{equation}

Since we are seeking isotropic configurations, the radial and tangential pressures must be equal, a condition that can be written as $G^r_r=G^\theta_\theta$. Using Eqs.~\eqref{metric} and \eqref{grr}, this condition leads to a differential equation for $\Phi^{(II)}$ \cite{Boonserm:2005ni}:
\begin{equation}
\left(8\pi r^5\rho_o+8\pi r^2R_i^3\rho_D-3r^3\right)
\left[\Phi''+(\Phi')^2\right]
+3r\left(-4\pi R_i^3\rho_D+r\right)\Phi'
-12\pi\rho_D R_i^3=0.
\end{equation}

The solution to this equation is given by
\begin{equation}
\label{phiii}
e^{\Phi^{(II)}}=\sqrt{B(r)}
\left(\bar C_1+\bar C_2 F(r)\right),
\end{equation}
where $\bar C_1$ and $\bar C_2$ are integration constants, and
\begin{equation}
\label{B}
B(R_i,\rho_o,\rho_D,r)
\equiv
1-\frac{8\pi}{3r}\rho_o\left(r^3+fR_i^3\right),
\end{equation}
\begin{equation}
\label{F}
F(R_i,\rho_o,\rho_D,r)
\equiv
\int_{R_i}^{r}
\frac{x^{5/2}\,dx}
{\left[
x-\frac{8\pi}{3}\rho_o(x^3+fR_i^3)
\right]^{3/2}},
\end{equation}
where the expression inside the parentheses on the left-hand side of the last two equations indicates the dependence on the parameter space $(R_i,\rho_o,\rho_D)$ and on the radial coordinate $r$.

We define the ratio between the dark matter and ordinary matter densities as $f=\rho_D/\rho_o$. The pressure can then be calculated from $G^r_r=8\pi p$, yielding
\footnote{To avoid clumsy notation, we shall avoid the explicit dependence on the parameters when possible.}
\begin{equation}
\label{PresionII}
p_o^{(II)}(R_i,\rho_o,\rho_D,r)
=
\frac{\bar C_2}
{4\pi\left(\bar C_1+\bar C_2F(r)\right)\sqrt{B(r)}}
-\rho_o.
\end{equation}
We can verify from Eqs.~\eqref{B} and \eqref{F} that, in the last expression, the functions $B(r)$ and $F(r)$ depend not only on the radial coordinate but also on the parameter space $(R_i,\rho_o,\rho_D)$.

As mentioned, the boundary conditions impose that the pressure in Region II must vanish at the surface of the star, namely, $p_o^{(II)}(r=R_e)=0$, which yields
\begin{equation}
\label{ValorDeC1}
\bar C_1
=
\bar C_2
\left[
\frac{1}{4\pi\rho_o\sqrt{B_e}}-F_e
\right],
\end{equation}
where $B_e\equiv B(r=R_e)$ and $F_e\equiv F(r=R_e)$.

Upon substitution of the relation between $\bar C_1$ and $\bar C_2$ in Eq.~\eqref{PresionII}, we obtain
\begin{equation}
\label{PresionIIA}
p_o^{(II)}(\rho_o,\rho_D,R_i,R_e,r)
=
\rho_o
\left[
\frac{\sqrt{B_e}}
{\sqrt{B(r)}
\left[
1+4\pi\rho_o\sqrt{B_e}\left(F(r)-F_e\right)
\right]}
-1
\right].
\end{equation}

Next, we impose the continuity of the pressure at $r=R_i$, namely,
\begin{equation}
\label{equalp}
p_T^{(I)}(r=R_i)
=
p_o^{(I)}(p_{TC},\rho_T,r=R_i)
=
p_o^{(II)}(\rho_o,\rho_D,R_i,R_e,r=R_i),
\end{equation}
where $p_o^{(I)}(R_i)$ is given by Eq.~\eqref{PresionTotalI1} evaluated at $r=R_i$.

As mentioned earlier, the parameter $p_{TC}$ is not free, but rather is determined by the boundary conditions of our system. Thus, from Eq.~\eqref{equalp}, it is straightforward to obtain the value of the central pressure in terms of the parameters of the model:
\begin{equation}
\label{ptc}
p_{TC}(\rho_o,\rho_D,R_i,R_e)
=
\rho_T
\frac{
\left[1-\sqrt{\beta_i}\right]
+\frac{1}{\rho_T}
\left[3-\sqrt{\beta_i}\right]
p_o^{(II)}(\rho_o,\rho_D,R_i,R_e,r=R_i)
}{
\left[3\sqrt{\beta_i}-1\right]
+\frac{3}{\rho_T}
\left[\sqrt{\beta_i}-1\right]
p_o^{(II)}(\rho_o,\rho_D,R_i,R_e,r=R_i)
}.
\end{equation}

In the limit $\rho_D=0$, $R_i=0$, this expression reduces to that of Schwarzschild's star, see Eq.~\eqref{pcs}.

Let us introduce some notation that will simplify the results. In the following we shall use
\begin{align}
C_i
&\equiv\frac{2M_i}{R_i},
\qquad
\mbox{with }
M_i=\frac{4\pi}{3}\rho_oR_i^3
\quad\rightarrow\quad
C_i=\frac{8\pi}{3}R_i^2\rho_o,
\label{Ci}\\
C_e
&\equiv\frac{2M_e}{R_e},
\qquad
\mbox{with }
M_e=\frac{4\pi}{3}\rho_oR_e^3
\quad\rightarrow\quad
C_e=\frac{8\pi}{3}R_e^2\rho_o.
\label{Ce}
\end{align}
It is straightforward to check that
\begin{equation}
\label{LimiteCiCe}
\lim_{R_e\to R_i}C_e=C_i.
\end{equation}

In this way, the structure of the configuration, which was determined by $(\rho_o,\rho_D,R_i,R_e)$, is now determined by $(\rho_o,f,C_i,C_e)$.

In terms of these quantities,
\begin{equation}
\label{RelacionBetaCi}
\beta(r)
=
1-C_i\left(\frac{r}{R_i}\right)^2(1+f)
\quad\Rightarrow\quad
\beta_i=1-C_i(1+f),
\end{equation}
and
\begin{align}
\label{Bi}
B_i&\equiv B(R_i)=1-(1+f)C_i=\beta_i,\\
\label{Be}
B_e&\equiv B(R_e)=1-(1+f\alpha^{-3})C_e,
\end{align}
where we have defined the ratio between the outer and inner radii as $\alpha=R_e/R_i$. Note that
\begin{equation}
\alpha^2=\frac{C_e}{C_i}.
\end{equation}

It is straightforward to check that
\begin{equation}
\lim_{f\to0}\beta_i
=
\lim_{f\to0}B_i
=
\lim_{f\to0,\,\alpha\to1}B_e
=
1-C_i.
\end{equation}

Defining $z=x/R_i$, it follows from Eq.~\eqref{F} that
\begin{equation}
\label{deff}
F(r)=R_i^2\varphi\left(f,C_i,r/R_i\right),
\end{equation}
with
\begin{equation}
\label{VarPhi}
\varphi\left(f,C_i,r/R_i\right)
=
\int_1^{r/R_i}
\frac{z^{5/2}\,dz}
{\left[z-C_i z^3-fC_i\right]^{3/2}}.
\end{equation}

Thus, it is direct to check that
\begin{equation}
\varphi_e
\equiv
\varphi(f,C_i,\alpha)
=
\varphi\left(f,C_i,r/R_i\right)\bigg|_{r=R_e}.
\end{equation}

Hence, from Eq.~\eqref{PresionIIA},
\begin{equation}
\label{PresionIIA1}
p_o^{(II)}(\rho_o,f,C_i,C_e,r)
=
\rho_o
\left[
\frac{\sqrt{B_e}}
{\sqrt{B(r)}
\left[
1+\frac{3}{2}\sqrt{B_e}\,C_i
\left(\varphi(r)-\varphi_e\right)
\right]}
-1
\right],
\end{equation}
which satisfies $p_o^{(II)}(R_e)=0$.

It follows from $B_e\geq0$ that
\begin{equation}
C_e\leq\frac{1}{1+f\alpha^{-3}}.
\end{equation}

Notice that this expression reduces to that of Schwarzschild's star if we set $\rho_D=0$ and $R_i=0$.

Using these parameters, and employing Eq.~\eqref{PresionIIA1}, we can write
\begin{equation}
\label{PresionIIA11}
p_{oi}^{(II)}
\equiv
p_o^{(II)}(\rho_o,f,C_i,C_e,r=R_i)
=
\rho_o
\left[
\frac{\sqrt{B_e}}
{\sqrt{B_i}
\left[
1-\frac{3}{2}\sqrt{B_e}\,C_i\varphi_e
\right]}
-1
\right].
\end{equation}

Thus, Eq.~\eqref{ptc} can be written as
\begin{equation}
\label{ptc2}
p_{TC}(\rho_o,f,C_i,C_e)
=
(1+f)\rho_o
\frac{
\left[1-\sqrt{\beta_i}\right](1+f)
+
\left[3-\sqrt{\beta_i}\right]
p_{oi}^{(II)}/\rho_o
}{
\left[3\sqrt{\beta_i}-1\right](1+f)
+
3\left[\sqrt{\beta_i}-1\right]
p_{oi}^{(II)}/\rho_o
}.
\end{equation}

It follows from $\beta_i\geq0$ that
\begin{equation}
C_i\leq\frac{1}{1+f}.
\end{equation}

\subsection{Temporal component of the metric tensor}

First, we evaluate the boundary conditions at the surface of the star, $r=R_e$. 
From Eqs.~\eqref{phiii} and \eqref{ValorDeC1}, we obtain
\begin{equation}
g_{tt}(r=R_e^-)
=
-\exp\left(2\Phi^{(II)}(r=R_e^-)\right)
=
-\left(\frac{\bar C_2}{4\pi\rho_o}\right)^2.
\end{equation}
On the other hand, since the exterior of the stellar distribution, denoted as 
Region III, corresponds to the vacuum, spherically symmetric Schwarzschild 
solution, we have
\begin{equation}
g_{tt}(r=R_e^+)
=
-\left(1-\frac{2M}{R_e}\right)
=
-\left[
1-C_i\left(\frac{f}{\alpha}+\alpha^2\right)
\right],
\end{equation}
where the parameter $M$ is given by Eq.~\eqref{MasaTotal}. Therefore, the 
continuity of $g_{tt}$ at $r=R_e$ yields
\begin{equation}
\label{ValorDeC2}
\bar C_2
=
4\pi\rho_o
\sqrt{
1-C_i\left(\frac{f}{\alpha}+\alpha^2\right)
}.
\end{equation}
Using Eq.~\eqref{Be}, this result can also be written as
\begin{equation}
\bar C_2=4\pi\rho_o\sqrt{B_e}.
\end{equation}

From Eqs.~\eqref{phiii}, \eqref{ValorDeC1}, and \eqref{deff}, the temporal
metric function in Region II can be written as
\begin{equation}
\label{phic2}
e^{\Phi^{(II)}(r)}
=
\frac{\sqrt{B(r)}\,\bar C_2}
{4\pi\rho_o\sqrt{B_e}}
\left[
1+\frac{3}{2}C_i\sqrt{B_e}
\left(\varphi(r)-\varphi_e\right)
\right].
\end{equation}
Using Eq.~\eqref{ValorDeC2}, this expression reduces to
\begin{equation}
e^{\Phi^{(II)}(r)}
=
\sqrt{B(r)}
\left[
1+\frac{3}{2}C_i\sqrt{B_e}
\left(\varphi(r)-\varphi_e\right)
\right].
\end{equation}

Now, we impose the continuity of the temporal component of the metric at the
inner boundary $r=R_i$. Since $\varphi(R_i)=0$, Eq.~\eqref{phic2} gives
\begin{equation}
\label{gttRi+}
\exp\left(2\Phi^{(II)}(r=R_i^+)\right)
=
B_i
\left[
1-\frac{3}{2}C_i\sqrt{B_e}\,\varphi_e
\right]^2.
\end{equation}
Equivalently, using Eqs.~\eqref{Bi} and \eqref{Be},
\begin{equation}
\exp\left(2\Phi^{(II)}(r=R_i^+)\right)
=
\left[1-C_i(1+f)\right]
\left[
1-\frac{3}{2}C_i
\sqrt{
1-C_i\left(\alpha^2+\frac{f}{\alpha}\right)
}
\,\varphi_e
\right]^2.
\end{equation}

On the other hand, from Eq.~\eqref{phipc1}, the temporal metric function in
Region I evaluated at $r=R_i$ is
\begin{equation}
\label{gttRi-}
\exp\left(2\Phi^{(I)}(r=R_i^-)\right)
=
\kappa_I
\left[
1-
\frac{3G+1}{3(G+1)}
\sqrt{\beta_i}
\right]^2,
\end{equation}
where
\begin{equation}
G=\frac{p_{TC}}{\rho_T},
\end{equation}
$p_{TC}$ is given by Eq.~\eqref{ptc2}, and $\beta_i$ is given by
Eq.~\eqref{Bi}.

Finally, imposing
\begin{equation}
\exp\left(2\Phi^{(I)}(R_i^-)\right)
=
\exp\left(2\Phi^{(II)}(R_i^+)\right)
\end{equation}
determines the remaining integration constant $\kappa_I$, yielding
\begin{equation}
\label{kappaI}
\kappa_I
=
\frac{
B_i
\left[
1-\dfrac{3}{2}C_i\sqrt{B_e}\,\varphi_e
\right]^2
}{
\left[
1-\dfrac{3G+1}{3(G+1)}
\sqrt{\beta_i}
\right]^2
}.
\end{equation}

Thus, the constants $\kappa_I$ and $\bar C_2$ determine the temporal components
of the metric in Regions I and II, respectively, and are fixed by the continuity
of the metric at $r=R_i$ and $r=R_e$.

The numerical behavior of the temporal metric component will be examined
later in the subsection \ref{AnalisisNumerico}, devoted to the numerical analysis. This allows
us to discuss the geometry together with the corresponding pressure
profiles for the same representative values of $f$ and $\alpha$, and to
illustrate explicitly the behavior of $g_{tt}$ across the core--envelope
interface.

\subsection{Analogous Buchdahl limit}

To investigate the critical compactness of the present configuration, we
consider the divergence of the central pressure. From
Eq.~\eqref{ptc2}, this occurs when its denominator vanishes, namely,
\begin{equation}
\label{BuchdahlDenominator}
\left(3\sqrt{B_i}-1\right)(1+f)
+3\left(\sqrt{B_i}-1\right)
\frac{p_{oi}^{(II)}}{\rho_o}=0.
\end{equation}
Solving Eq.~\eqref{BuchdahlDenominator} for
$p_{oi}^{(II)}/\rho_o$, we obtain
\begin{equation}
\label{BuchdahlPressureCondition}
\frac{p_{oi}^{(II)}}{\rho_o}
=
\frac{1+f}{3}
\frac{1-3\sqrt{B_i}}{\sqrt{B_i}-1}.
\end{equation}

Substituting Eq.~\eqref{PresionIIA11} into
Eq.~\eqref{BuchdahlPressureCondition}, and using the definitions of
$B_i$ and $B_e$, we obtain
\begin{equation}
\label{BeBuchdahl}
B_e=
\left[
\frac{
\sqrt{B_i}
\left[
f\left(3\sqrt{B_i}-1\right)+2
\right]
}{
\frac{3}{2}C_i\varphi_e\sqrt{B_i}
\left[
f\left(3\sqrt{B_i}-1\right)+2
\right]
-3\left(\sqrt{B_i}-1\right)
}
\right]^2.
\end{equation}
Since
\begin{equation}
B_e=1-C_e\left(1+f\alpha^{-3}\right),
\end{equation}
the critical value of $C_e$ is therefore given by
\begin{equation}
\label{Buchdahl}
C_e=
\frac{1}{1+f\alpha^{-3}}
\left\{
1-
\left[
\frac{
\sqrt{B_i}
\left[
f\left(3\sqrt{B_i}-1\right)+2
\right]
}{
\frac{3}{2}C_i\varphi_e\sqrt{B_i}
\left[
f\left(3\sqrt{B_i}-1\right)+2
\right]
-3\left(\sqrt{B_i}-1\right)
}
\right]^2
\right\},
\end{equation}
where
\begin{equation}
B_i=1-(1+f)C_i,
\qquad
\varphi_e=\varphi(f,C_i,\alpha).
\end{equation}

It is useful to distinguish $C_e$, which is defined in terms of the
ordinary-matter contribution, from the total compactness of the
configuration. Using the total mass, the latter is
\begin{equation}
\label{TotalCompactnessBuchdahl}
C\equiv\frac{2M}{R_e}
=
C_e\left(1+f\alpha^{-3}\right)
=
C_i\left(\alpha^2+\frac{f}{\alpha}\right).
\end{equation}
Thus, Eq.~\eqref{Buchdahl} also determines the critical total compactness
at which the central pressure diverges.

As a consistency check, let us consider the one-fluid limit. When
$f\rightarrow0$ and $\alpha\rightarrow1$, one has
$\varphi_e\rightarrow0$, $C_e\rightarrow C_i$, and
$B_i\rightarrow1-C_i$. Equation~\eqref{Buchdahl} then reduces to
\begin{equation}
\label{BuchdahlSchwarzschild}
C_i=
1-
\frac{
4(1-C_i)
}{
9\left(\sqrt{1-C_i}-1\right)^2
}.
\end{equation}
The nontrivial physical solution of
Eq.~\eqref{BuchdahlSchwarzschild} is
\begin{equation}
C_i=\frac{8}{9},
\end{equation}
which reproduces the standard Buchdahl limit of the Schwarzschild
constant-density star.

Equation~\eqref{Buchdahl}, together with the relation
\begin{equation}
C_e=\alpha^2C_i,
\end{equation}
defines a highly nonlinear equation for the parameters
$(\alpha,C_i,f)$. In general, this equation cannot be solved analytically,
and therefore we proceed with a numerical analysis.

For fixed values of $\alpha$, we first identify the ranges of the
parameters $C_i$ and $f$ for which the metric functions and the integral
$\varphi$ are real and well defined. In particular, we require $B(r)$,
given by Eq.~\eqref{B}, as well as the argument in the denominator of the
integral in Eq.~\eqref{VarPhi}, to remain positive throughout the
corresponding region. Within this parameter space, we then analyze the
central pressure $p_{TC}$ given by Eq.~\eqref{ptc2} and the critical
condition given by Eq.~\eqref{Buchdahl}.

\begin{figure}[h]
\begin{center}
\includegraphics[width=3in]{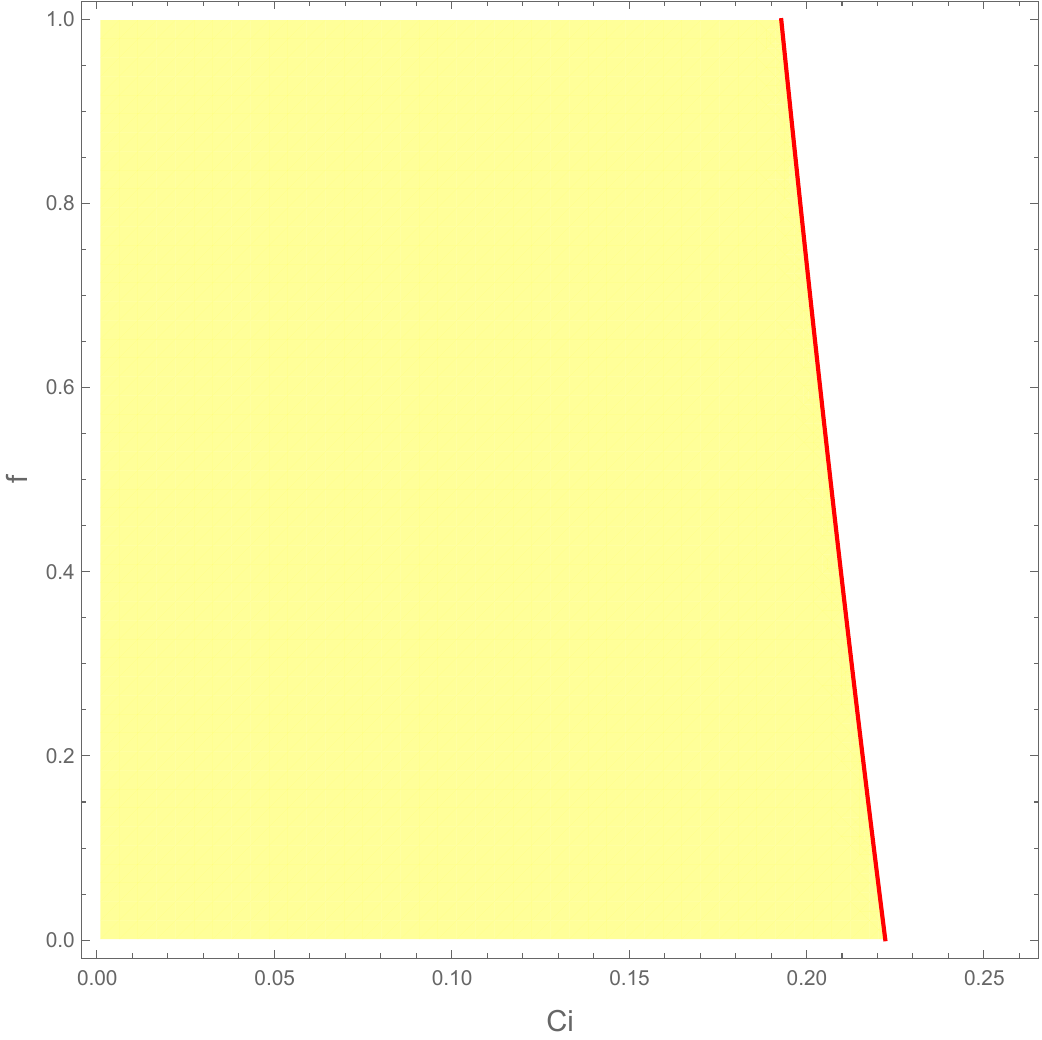}
\includegraphics[width=3in]{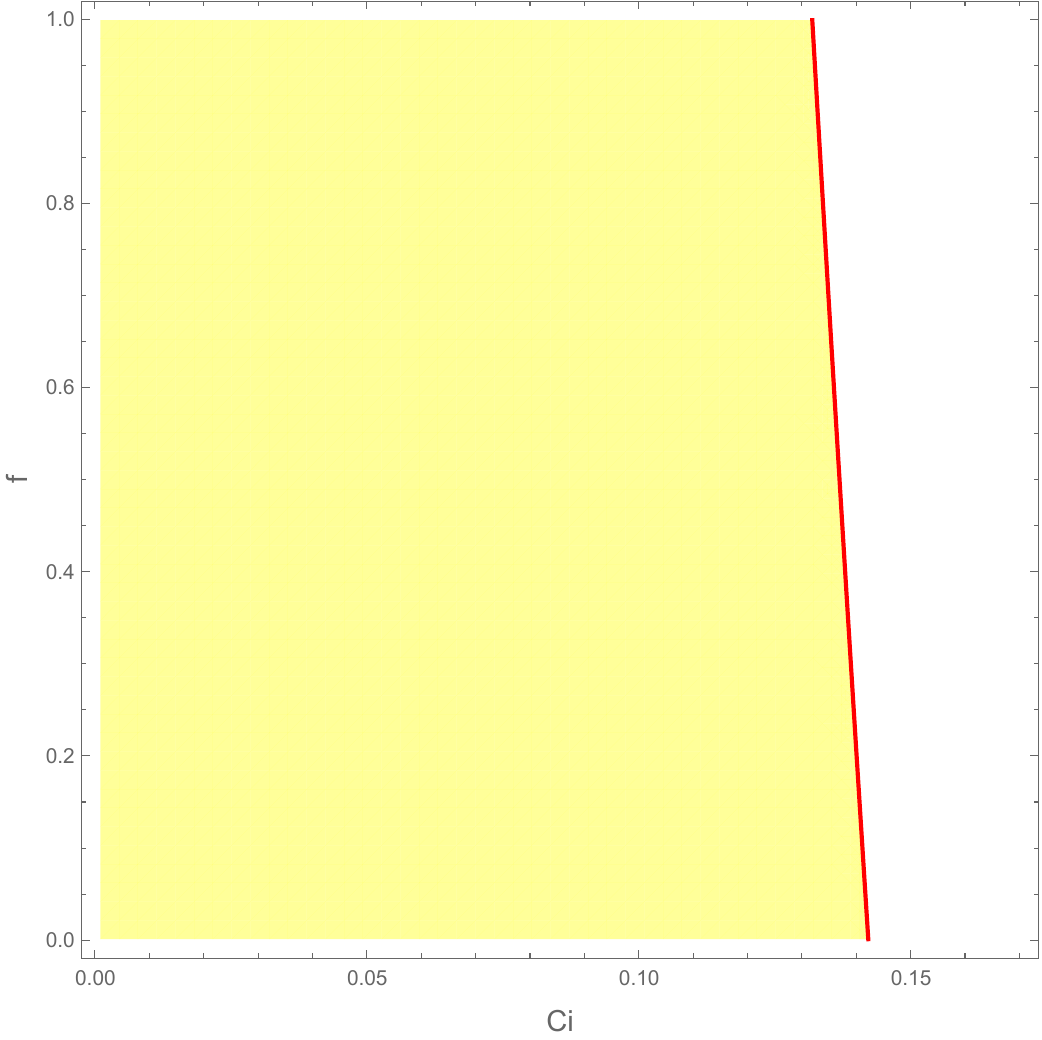}
\includegraphics[width=3in]{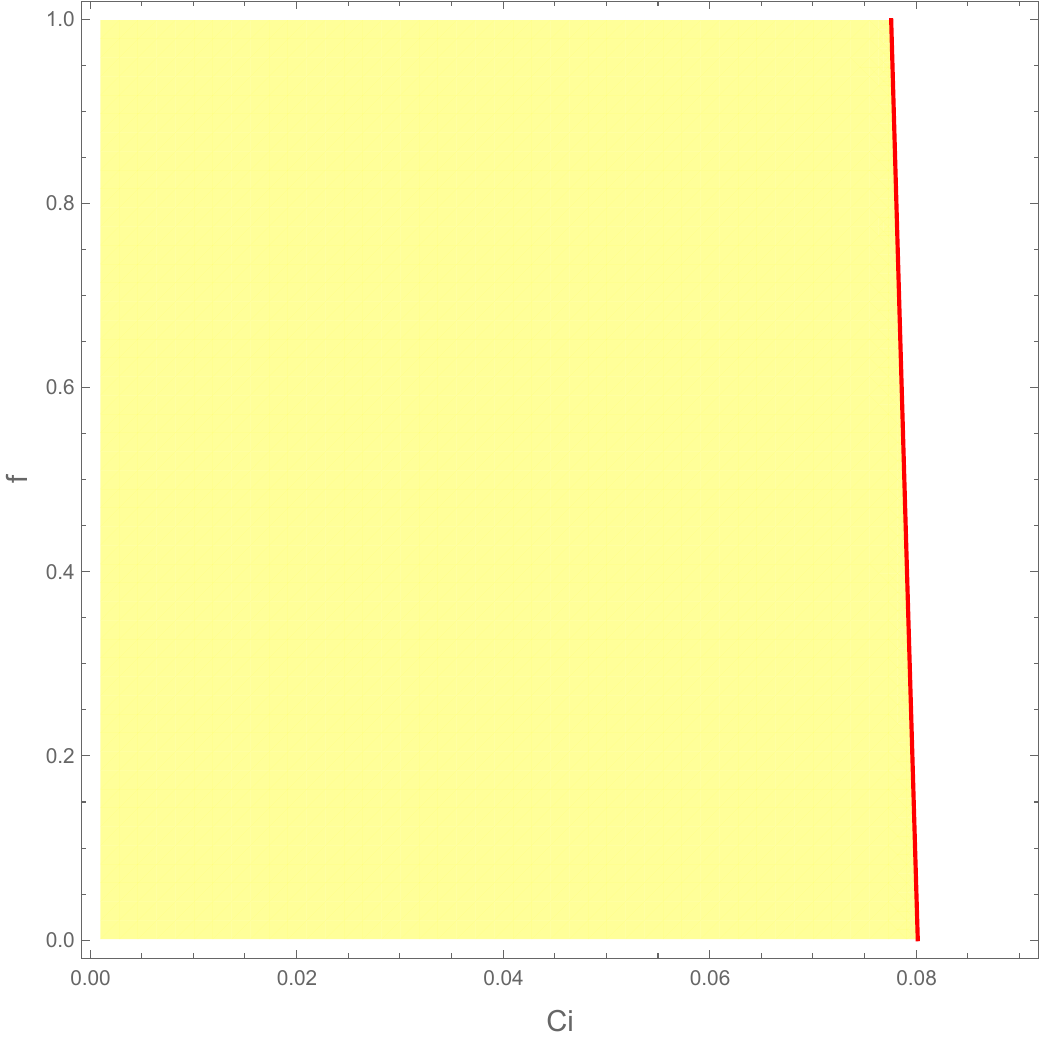}
\caption{Values of $(C_i,f)$ for $\alpha=2$, $2.5$, and $3.33$ in the
first, second, and third panels, respectively. The yellow region
corresponds to configurations satisfying the static admissibility
conditions considered here, including a finite and positive central
pressure and real metric functions. The solid red curve corresponds to
the pairs $(C_i,f)$ for which the central pressure diverges and therefore
represents the Buchdahl-like critical boundary of the present
configuration. This critical curve should not be interpreted as a
dynamical stability boundary.}
\label{FigBuchda}
\end{center}
\end{figure}

In Fig.~\ref{FigBuchda}, for fixed values of $\alpha$, the yellow region
represents the parameter ranges for which the configurations are
physically admissible under the static conditions considered here, in
particular, with a finite and positive central pressure and real metric
functions. The solid red curve corresponds to the pairs $(C_i,f)$ for
which the denominator of Eq.~\eqref{ptc2} vanishes and, consequently, the
central pressure diverges. Therefore, this curve represents a
Buchdahl-like critical boundary for the present two-fluid configuration.
We emphasize that this boundary is associated with the divergence of the
central pressure and should not be interpreted as a dynamical stability
boundary.

The approximately linear behavior of the critical curves displayed in
Fig.~\ref{FigBuchda} can be understood by examining the integral
$\varphi_e$. From Eq.~\eqref{VarPhi}, the denominator of its integrand
contains the cubic polynomial
\begin{equation}
\label{PolynomialVarphi}
P(z)=z-C_i z^3-fC_i.
\end{equation}
For $\varphi_e$ to remain real throughout the integration interval
$1\leq z\leq\alpha$, the polynomial $P(z)$ must remain positive over this
interval. The numerical analysis indicates that, along the physically
relevant critical branch, the largest positive root of $P(z)$ approaches
the upper integration limit $z=\alpha$. As this root approaches $\alpha$
from above, the denominator of the integrand becomes increasingly small
near the upper integration limit and $\varphi_e$ grows rapidly.
Consequently, the location of the critical branch can be approximately
described by $P(\alpha)\simeq0$. Using Eq.~\eqref{PolynomialVarphi}, this condition gives $\alpha-C_i\alpha^3-fC_i\simeq0$, and therefore
\begin{equation}
\label{ApproxCriticalCi}
C_i\simeq
\frac{\alpha}{\alpha^3+f}
=
\frac{1}{\alpha^2}
\frac{1}{1+f/\alpha^3}.
\end{equation}

Equation~\eqref{ApproxCriticalCi} provides a simple interpretation of the
nearly linear shape of the critical curves in the $(C_i,f)$ plane. In
the regime $f\ll\alpha^3$, it can be expanded as $C_i\simeq
\frac{1}{\alpha^2}
-\frac{f}{\alpha^5}
+\mathcal{O}\left(\frac{f^2}{\alpha^8}\right)$, or, to first order,
\begin{equation}
\label{ApproxLinearCritical}
f\simeq\alpha^3-\alpha^5 C_i.
\end{equation}
Thus, for fixed $\alpha$, the critical branch is approximately a straight
line in the $(C_i,f)$ plane, with slope
$df/dC_i\simeq-\alpha^5$. For the values used in Fig.~\ref{FigBuchda}, namely $\alpha=2$, $2.5$,
and $3.33$, with $0<f<1$, the expansion parameter satisfies
$f/\alpha^3<0.125$, $0.064$, and $0.027$, respectively. The linear
approximation therefore improves as $\alpha$ increases. Conversely, as
$f$ increases for fixed $\alpha$, higher-order corrections become more
relevant and the red critical branch appears as a slightly deformed
straight line. This effect is more noticeable for $\alpha=2$, whereas
for $\alpha=2.5$ and especially $\alpha=3.33$ the branch remains closer
to the linear approximation.

It is important to stress that the procedure described above provides
only an analytical interpretation of the shape and location of the
numerically determined critical branch. In particular,
$P(\alpha)\simeq0$ describes the regime in which the largest root
approaches the upper integration boundary and $\varphi_e$ becomes large;
it is not an independent exact condition for the divergence of the
central pressure. The Buchdahl-like critical configurations remain
defined by Eq.~\eqref{Buchdahl}.

\subsection{Numerical behavior analysis} \label{AnalisisNumerico}

\begin{figure}[h]
  \begin{center}
      \includegraphics[width=3in]{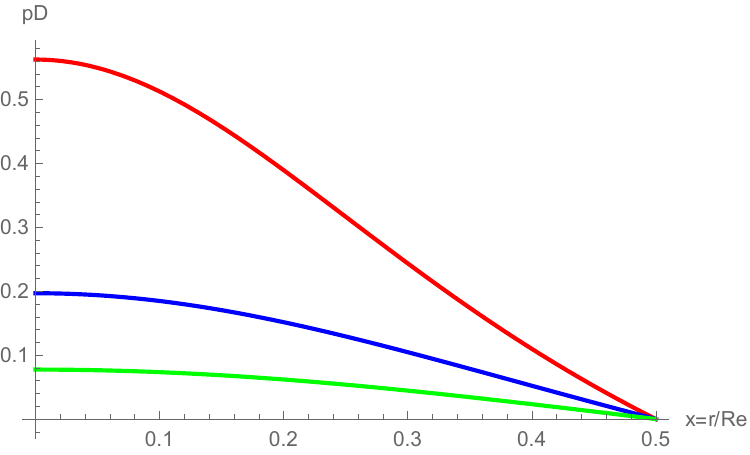}
      \includegraphics[width=3in]{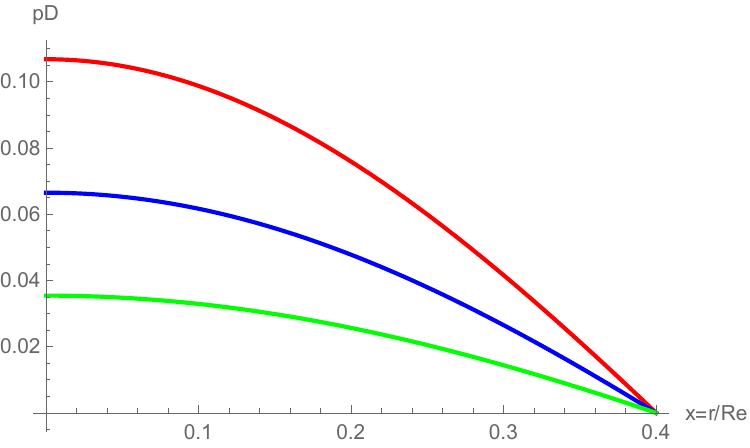}
      \includegraphics[width=3in]{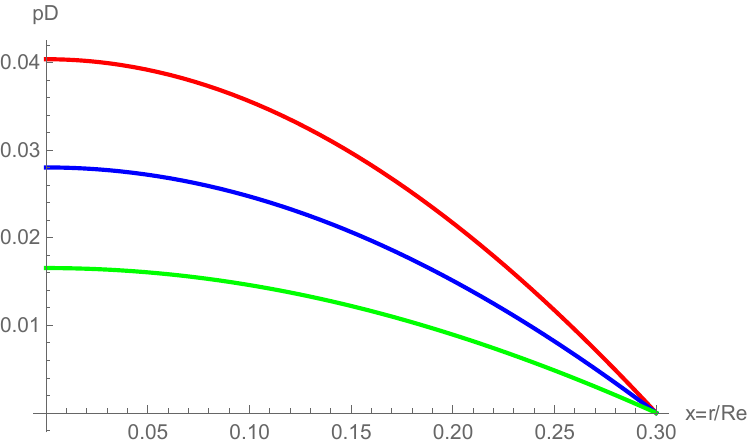}
        \caption{$p_D(x)$ with $x = r / R_e$ for $\alpha^{-1} = R_i / R_e = 0.5, 0.4,$ and $0.3$ in the first, second, and third panels, respectively, taking $f = \rho_D / \rho_0 = 0.972$ (red), $0.75$ (blue), and $0.5$ (green). }
\label{FigPDvariandoF}
  \end{center}
\end{figure}

We begin by analyzing the numerical behavior of the dark matter pressure in Region I, as given by Eqs. $\eqref{PresionDark}$ and $\eqref{kappaTregionI}$. As noted in the previous section, the central pressure $p_{TC}$ appearing in these expressions is not a free parameter; rather, it must be determined from the physical constraints imposed by the system, leading to Eq. $\eqref{ptc}$. Consequently, Figure \ref{FigPDvariandoF} depends on the parameters $f = \rho_D / \rho_0$ and $\alpha = R_e / R_i$.

\begin{figure}[h]
  \begin{center}
      \includegraphics[width=3.25in]{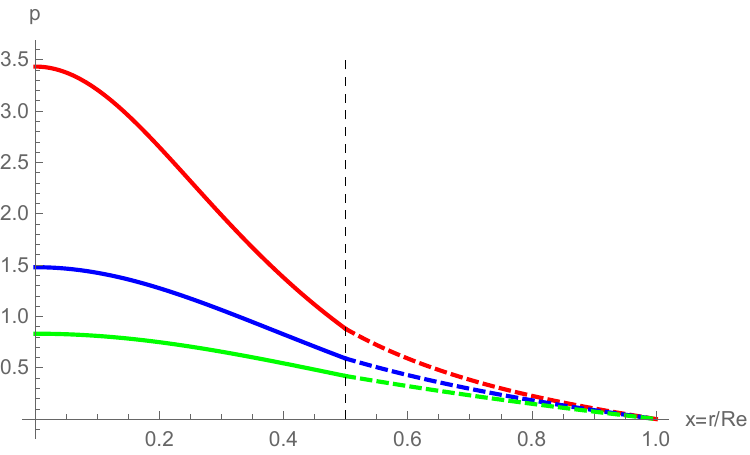}
      \includegraphics[width=3.25in]{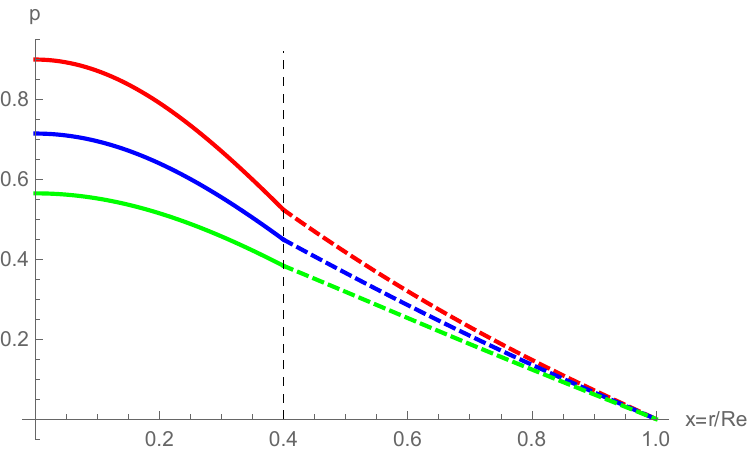}
      \includegraphics[width=3.25in]{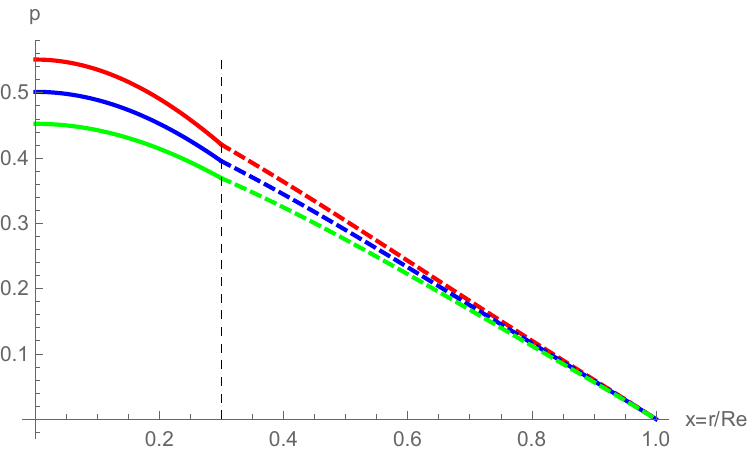}
        \caption{$p_T(x)$ with $x = r / R_e$, for $\alpha^{-1} = R_i / R_e = 0.5, 0.4,$ and $0.3$ in the first, second, and third panels, respectively. The curves correspond to $f = \rho_D / \rho_0 = 0.972$ (red), $0.75$ (blue), and $0.5$ (green). The solid line represents Region I, $x \in [0, R_i / R_e]$, while the dashed line represents Region II, $x \in [R_i / R_e, 1]$. }\label{FigPTvariandoF}
  \end{center}
\end{figure}

Also taking into account the physical constraints discussed in Figure \ref{FigPTvariandoF}, we display the numerical behavior of the total pressure: in Region I it is represented by the solid line, while in Region II it is given by the dashed line.

To complement the analysis of the matter sector, we now examine the
temporal metric component throughout the stellar interior. We use the
same representative values $f=0.972,\,0.75,$ and $0.5$ considered in
the pressure profiles, for $\alpha=2,\,2.5,$ and $3.33$. In Region I
and Region II, $g_{tt}$ is evaluated from Eqs.~\eqref{phipc1} and
\eqref{phiii}, respectively.

\begin{figure}[h]
  \begin{center}
      \includegraphics[width=3.5in]{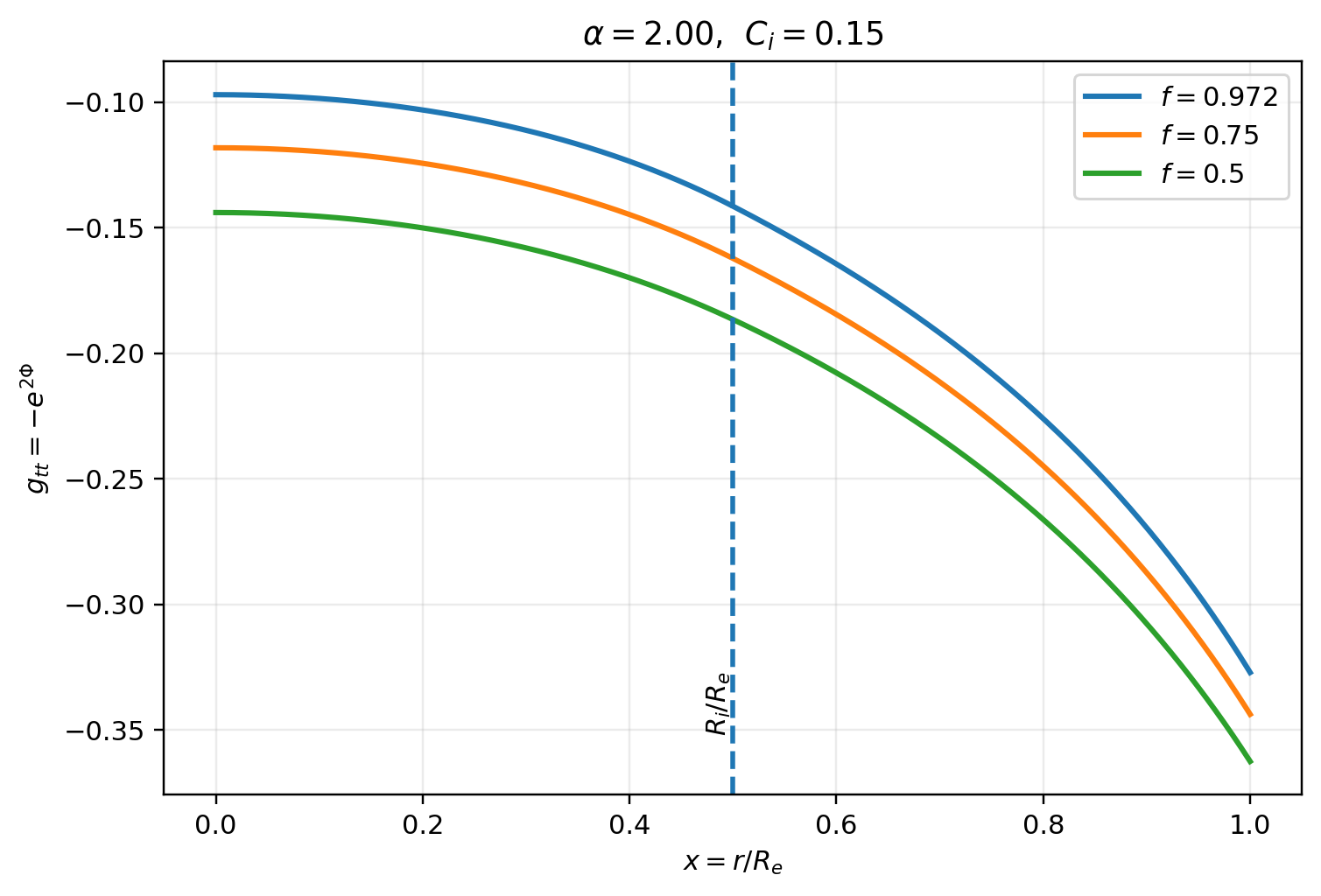}
      \includegraphics[width=3.5in]{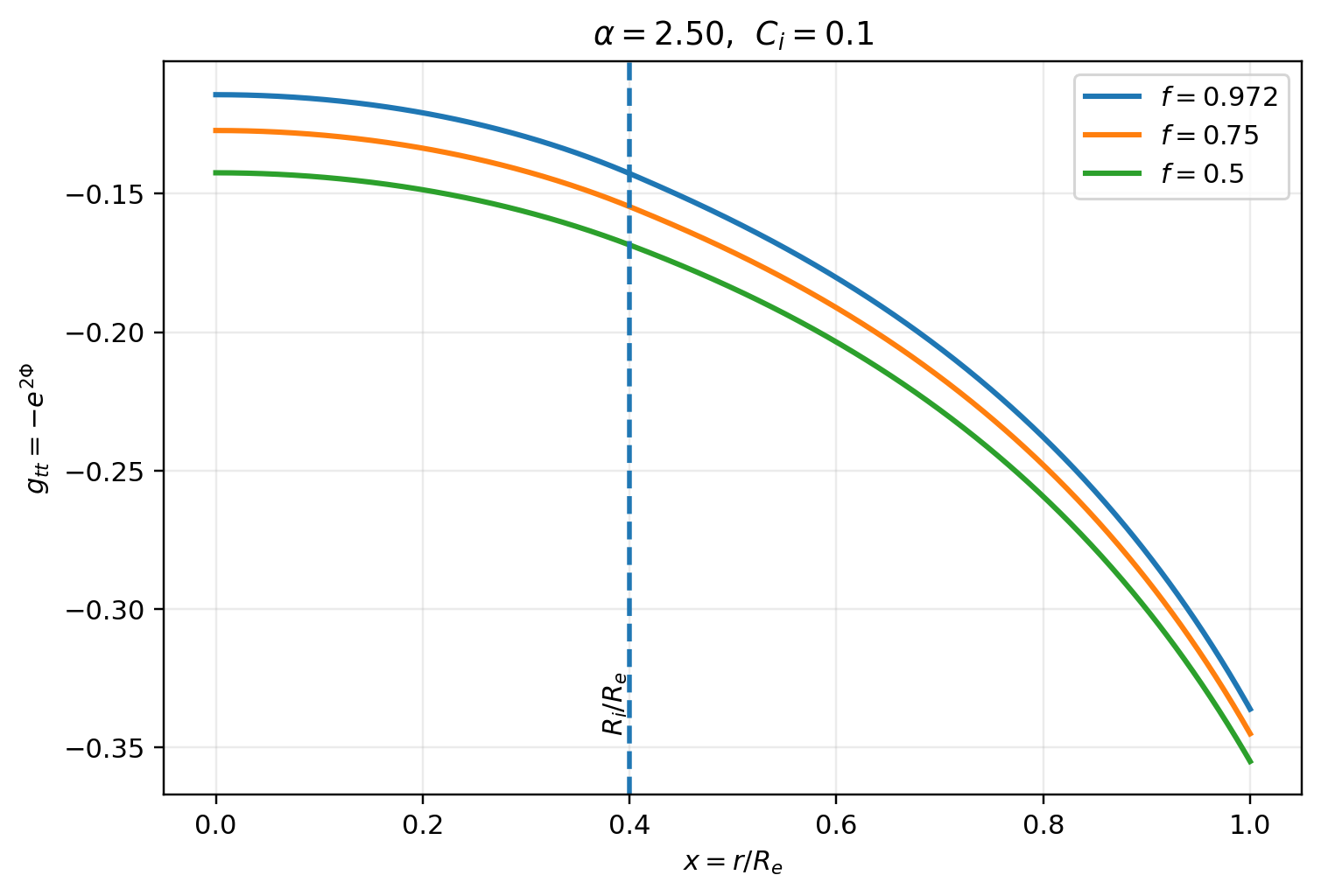}
      \includegraphics[width=3.5in]{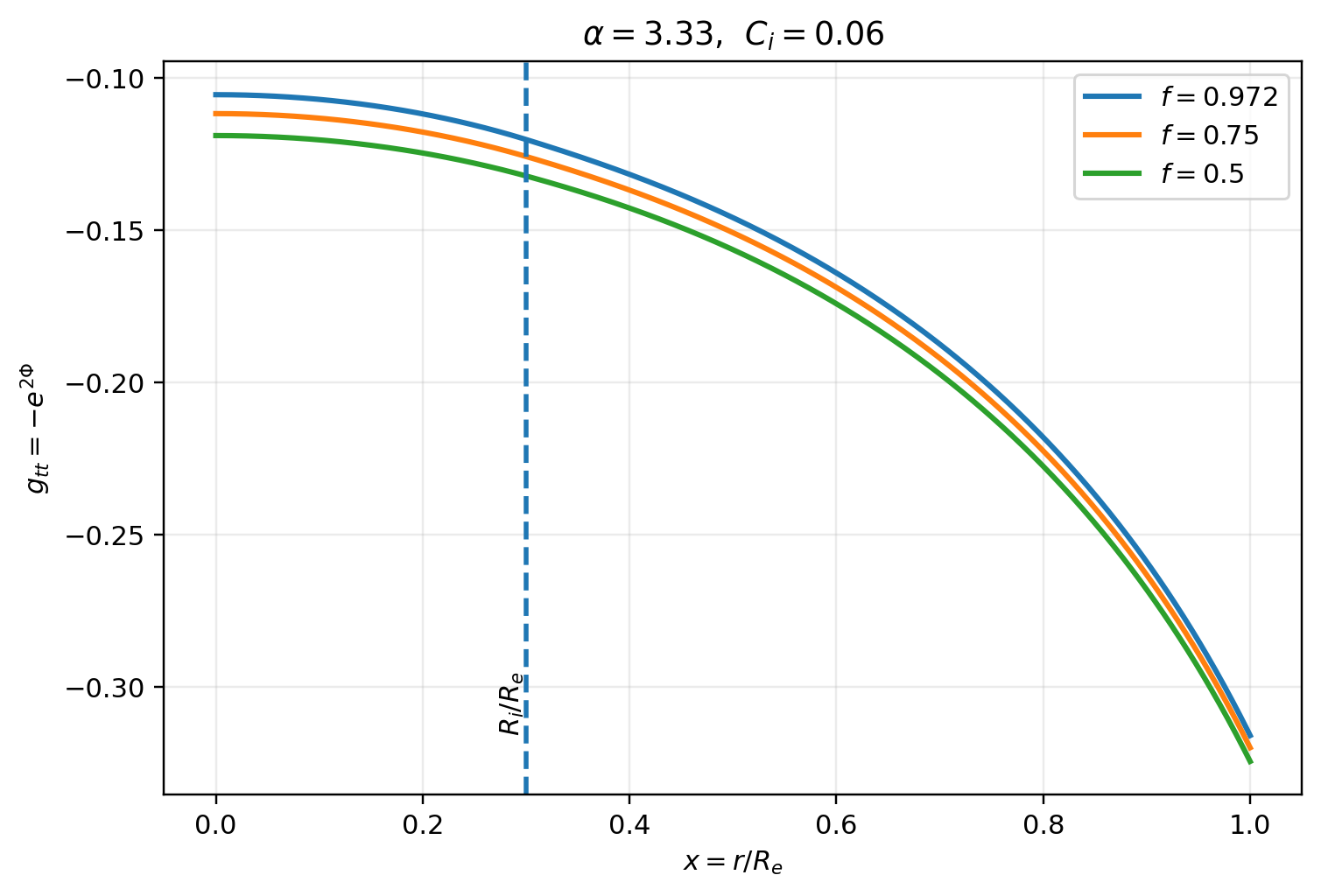}
      \caption{Temporal metric component $g_{tt}=-e^{2\Phi}$ as a
      function of $x=r/R_e$, for $\alpha=2$, $2.5$, and $3.33$
      in the first, second, and third panels, respectively. The curves
      correspond to $f=\rho_D/\rho_0=0.972$ (red), $0.75$ (blue),
      and $0.5$ (green). The vertical dashed line indicates the
      core--envelope interface $x=R_i/R_e=\alpha^{-1}$.}
      \label{FigGtt}
  \end{center}
\end{figure}

Figure~\ref{FigGtt} shows that $g_{tt}$ remains continuous across the
core--envelope interface, consistently with the matching conditions.
Although the dark component is confined to Region I, its contribution
to the gravitational field also affects the geometry of the
ordinary-matter envelope through the matching conditions.

\section{$M$--$R$ diagram}

The total mass of the configuration is given by
\begin{equation}
\label{MassMR}
M=\frac{4\pi}{3}\left(\rho_D R_i^3+\rho_o R_e^3\right).
\end{equation}
Using the definitions
\begin{equation}
f=\frac{\rho_D}{\rho_o},
\qquad
\alpha=\frac{R_e}{R_i},
\end{equation}
Eq.~\eqref{MassMR} can be written as
\begin{equation}
\label{MassMR2}
M=\frac{4\pi}{3}\rho_o R_e^3
\left(1+f\alpha^{-3}\right).
\end{equation}

The total compactness of the configuration is therefore
\begin{equation}
\label{CompactnessMR2}
C\equiv\frac{2M}{R_e}
=C_i\left(\alpha^2+\frac{f}{\alpha}\right)
=\alpha^2 C_i\left(1+f\alpha^{-3}\right).
\end{equation}
Notice that this is precisely the total compactness introduced in the
discussion of the Buchdahl-like critical condition.

For fixed values of $(\rho_o,f,\alpha)$, Eq.~\eqref{MassMR2} defines the
mass--radius relation
\begin{equation}
\label{MRcurve}
M(R_e)=
\frac{4\pi}{3}\rho_o
\left(1+f\alpha^{-3}\right)R_e^3.
\end{equation}
Therefore, for fixed $(\rho_o,f,\alpha)$, the mass increases cubically
with the stellar radius. The contribution of the second fluid enters
through the factor $f\alpha^{-3}$, reflecting the fact that this
component is restricted to the inner region $R_i=R_e/\alpha$.

It is convenient to introduce the dimensionless quantities
\begin{equation}
\mathcal{R}=\sqrt{\frac{8\pi\rho_o}{3}}\,R_e,
\qquad
\mathcal{M}=\sqrt{\frac{8\pi\rho_o}{3}}\,M.
\end{equation}
Equation~\eqref{MRcurve} then becomes
\begin{equation}
\label{DimensionlessMR}
\mathcal{M}
=
\frac{1}{2}
\left(1+f\alpha^{-3}\right)
\mathcal{R}^3.
\end{equation}
This representation allows configurations with different dark-matter
fractions and relative core sizes to be compared without introducing an
arbitrary dimensional scale.

For each fixed pair $(f,\alpha)$, we construct the curve by varying
$R_e$, or equivalently $\mathcal{R}$, while keeping $\rho_o$, $f$, and
$\alpha$ fixed. Since
\begin{equation}
C_i=\frac{8\pi}{3}\rho_o R_i^2
=\frac{8\pi}{3}\rho_o\frac{R_e^2}{\alpha^2}
=\frac{\mathcal{R}^2}{\alpha^2},
\end{equation}
increasing the stellar radius also increases $C_i$. The sequence is
continued only through the physically admissible parameter region
identified in the previous sections and is terminated when the
denominator of the central pressure in Eq.~\eqref{ptc2} vanishes. The
endpoint therefore corresponds to the Buchdahl-like critical
compactness associated with $p_{TC}\rightarrow\infty$ within the
present model.

The present analysis concerns the existence and static physical
admissibility of the configurations and does not establish their
dynamical stability. In particular, the critical curve obtained from
the divergence of the central pressure defines a Buchdahl-like critical
compactness within the present construction, rather than a
marginal-stability curve. This distinction is especially relevant for
two-fluid relativistic stars. As shown in \cite{Caballero2024}, radial
stability in systems composed of two perfect fluids requires the
analysis of the coupled radial perturbation equations, with stability
determined by the sign of the squared frequency of the fundamental
radial mode. Consequently, configurations lying within the physically
admissible region identified here are not necessarily dynamically
stable. A radial-mode analysis of the present analytical configurations
would therefore provide a natural extension of this work and would
allow the stability boundary to be compared directly with the critical
compactness obtained here.

\begin{figure}[h]
\begin{center}
\includegraphics[width=4.5in]{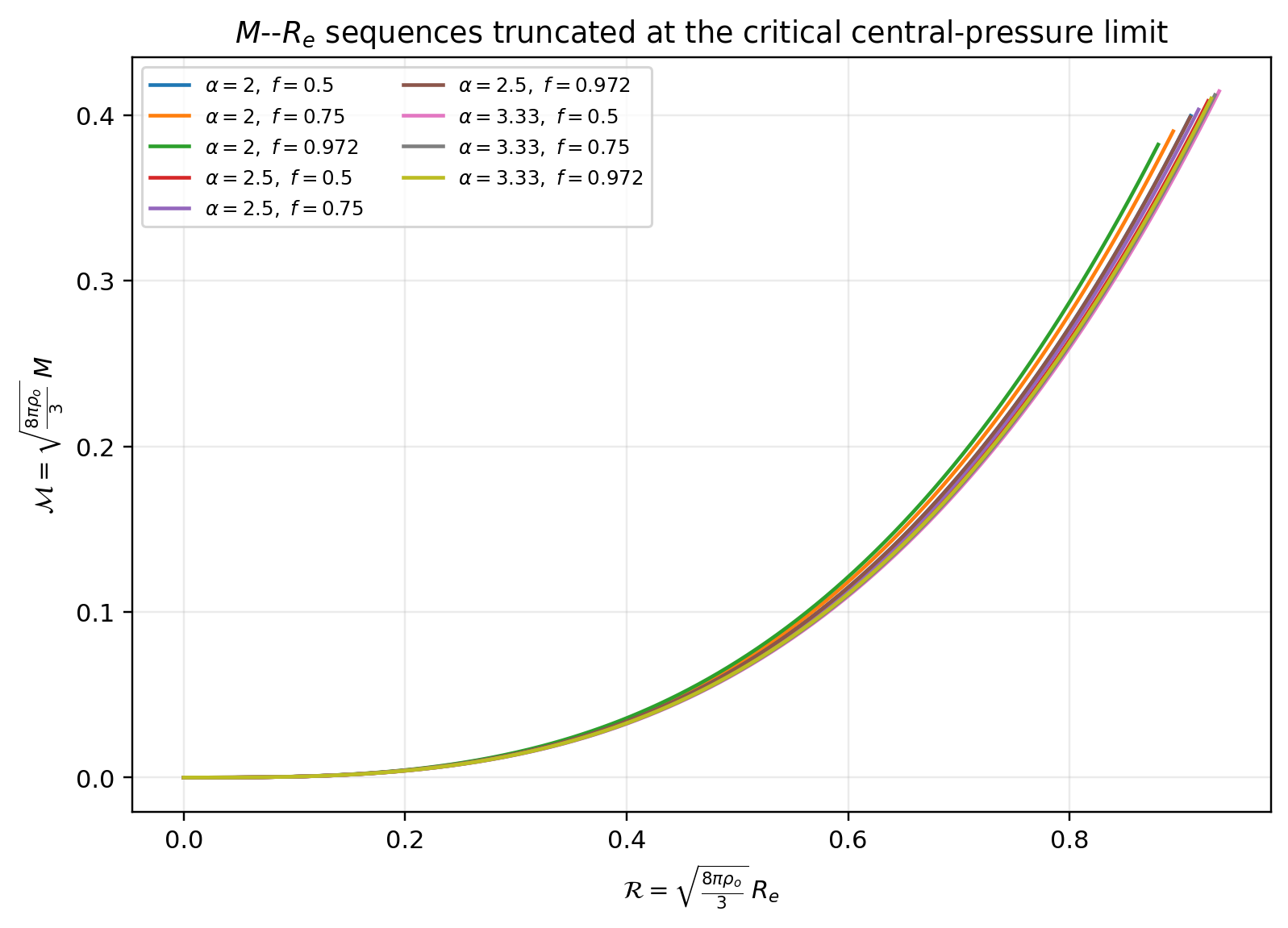}
\caption{Dimensionless mass--radius relations for different values of
the dark-matter fraction $f$ and the ratio $\alpha=R_e/R_i$. Each curve
is obtained by keeping $(\rho_o,f,\alpha)$ fixed and varying the stellar
radius. The curves are terminated at the critical configurations for
which the central pressure diverges, $p_{TC}\rightarrow\infty$,
corresponding to the Buchdahl-like critical condition derived in
Eq.~\eqref{Buchdahl}. These endpoints should not be interpreted as
marginal-stability configurations.}
\label{FigMR}
\end{center}
\end{figure}

Several features can be inferred from Fig.~\ref{FigMR}. For a fixed
radius and fixed ordinary-matter density, Eq.~\eqref{MRcurve} shows that
the presence of the second fluid increases the total mass through the
factor $1+f\alpha^{-3}$. This effect is more pronounced when the mixed
core occupies a larger fraction of the star. Indeed, for increasing
$\alpha$, the core radius $R_i=R_e/\alpha$ becomes smaller relative to
the stellar radius, and the contribution of the dark component to the
total mass is suppressed as $\alpha^{-3}$.

The numerical results also indicate that, for fixed $\alpha$, increasing
$f$ shifts the critical endpoint of the sequence. In particular, the
critical value of $C_i$ at which the central pressure diverges decreases
as the relative dark-matter density increases. Thus, within the present
model, a larger dark-matter contribution modifies not only the mass of
the configuration but also the range of parameters accessible before
the critical central-pressure condition is reached. This behavior is a
property of the static solutions and should not by itself be interpreted
as a statement about dynamical stability, which requires an independent
radial-perturbation analysis.

Equation~\eqref{CompactnessMR2} also shows that different combinations of
$(C_i,f,\alpha)$ can produce the same total compactness $C$. Therefore,
configurations with the same value of $M/R_e$ may nevertheless possess
different internal structures, characterized by different dark-matter
fractions and different relative core sizes. This degeneracy is a
consequence of the additional degrees of freedom introduced by the
two-component structure and is absent in the standard one-fluid
Schwarzschild constant-density configuration.

It is important to emphasize that the curves displayed in
Fig.~\ref{FigMR} should not be interpreted as realistic neutron-star
sequences associated with a microscopic equation of state. Rather, they
provide an analytically controlled representation of how a centrally
concentrated, gravitationally coupled second fluid modifies the
mass--radius relation and the critical compactness of the Schwarzschild
constant-density configuration. In particular, the analytical character
of the model makes it possible to trace these modifications directly to
the relative density $f$ and the relative core size $\alpha$, while
retaining the standard Schwarzschild configuration as the corresponding
one-fluid limit.

\subsection{Connection between internal structure and global properties}

The numerical results allow us to connect the internal structure of the
configurations with their global properties. Figures~\ref{FigPDvariandoF}--\ref{FigGtt}
use the representative values $\alpha=2,\,2.5,\,3.33$ and
$f=0.972,\,0.75,\,0.5$, while Fig.~\ref{FigBuchda} explores $0<f<1$
for the same values of $\alpha$. This makes it possible to relate the
pressure and metric profiles directly to the critical boundary in the
$(C_i,f)$ plane. In particular, for fixed $\alpha$, the critical value
of $C_i$ decreases as $f$ increases, while the pressure profiles show
how different dark-matter fractions are reflected in the internal matter
distribution. The influence of the second component is also reflected in the geometry.
As shown in Fig.~\ref{FigGtt}, $g_{tt}$ remains continuous across
$R_i$, while its radial behavior depends on the matter parameters.
Although the dark component is confined to the core, its gravitational
contribution enters the matching conditions and is therefore reflected
also in the geometry of the ordinary-matter envelope. Finally, Fig.~\ref{FigMR} connects these internal properties with the
global mass--radius sequences. As $C_i$ increases along a sequence with
fixed $(\rho_o,f,\alpha)$, the configuration approaches the critical
boundary associated with $p_{TC}\rightarrow\infty$. Taken together, the
figures indicate that the fraction and spatial extent of the second
component are associated with changes in the pressure profiles, geometry,
and critical global properties. Moreover, configurations with the same
global compactness may have different internal matter distributions,
suggesting that $M/R_e$ alone does not uniquely characterize the internal
structure within the present model.

\section{Conclusions}

In this work, we have constructed an analytical two-fluid extension of the Schwarzschild constant-density star in which two independently conserved incompressible perfect fluids coexist in an inner core, while only one component extends throughout the outer envelope. The two fluids exchange neither matter nor energy and interact only through the common spacetime geometry. This mixed-core/single-fluid-envelope structure provides an analytically controlled setting for isolating the gravitational effects of a centrally concentrated second component, which in the present work is interpreted as dark matter.

Despite the two-fluid and core--envelope structure, the model remains largely analytically tractable. We obtained explicit expressions for the pressures and metric functions in both regions and determined the central pressure and integration constants from the boundary and matching conditions. In particular, the central pressure is fixed in terms of the relative density $f=\rho_D/\rho_o$, the relative size of the mixed core, and the compactness parameters. The standard Schwarzschild constant-density solution is recovered in the corresponding one-fluid limit.

The analytical character of the construction also allowed us to derive a Buchdahl-like critical compactness associated with the divergence of the central pressure. This critical value depends explicitly on both the relative density of the second fluid and the relative size of the mixed core, while the standard Schwarzschild value $2M/R=8/9$ is recovered when the second component vanishes and the distinction between the core and envelope is removed. The numerical analysis identifies the physically admissible parameter region under the conditions considered here and illustrates how the additional component modifies the pressure profiles and the critical configurations.

The mass--radius analysis provides a complementary characterization of these effects. Besides modifying the total mass and the critical endpoint of the sequences, the additional degrees of freedom associated with $f$ and the core size allow configurations with the same global compactness to possess different internal matter distributions. The construction therefore goes beyond a simple modification of the total mass: it provides an explicit framework in which changes in the internal two-fluid structure can be related directly to changes in the pressure distribution, mass--radius relation, and Buchdahl-like critical compactness. More realistic two-fluid models indicate that dark matter can affect global properties of neutron stars, including their compactness and tidal deformability \cite{Liu:2024}; extending the present analytical framework to quantities such as tidal deformability would thus be a natural direction for future work. 

Although a more detailed assessment of the applicability of the present construction to realistic neutron stars requires the inclusion of microscopic equations of state and a dedicated stability analysis, which are beyond the scope of this work, its analytical character may provide
useful insight into how a gravitationally coupled second component affects stellar structure. In this sense, the model can serve as an analytical benchmark for identifying qualitative trends that can subsequently be tested in more realistic dark-matter-admixed neutron-star models.

The present analysis concerns the existence and static physical admissibility of the configurations and does not establish their dynamical stability. In particular, the critical curve obtained from the divergence of the central pressure should not be interpreted as a marginal-stability curve. As shown in \cite{Caballero2024}, radial stability in relativistic two-perfect-fluid systems requires an analysis of the coupled radial perturbations and the corresponding fundamental mode. Applying this formalism to the configurations obtained here would allow the actual stability boundary to be compared with the Buchdahl-like critical compactness derived in this work. Other extensions could include more realistic equations of state and limiting configurations such as gravastar-like regimes.

Overall, the present model provides an analytically controlled extension of the Schwarzschild constant-density star in which the effects of an independently conserved, centrally concentrated second component can be followed explicitly from the internal pressure distribution and core--envelope matching to the global mass--radius relation and critical compactness. Its main role is therefore not to provide a microscopic description of a realistic dark-matter-admixed neutron star, but to offer a tractable analytical benchmark for identifying and disentangling the gravitational effects introduced by an additional matter component.

\appendix

\section{A brief review of the Schwarzschild interior solution}
\label{ApendiceEstrellaSch}

Starting from the metric \eqref{MetricaRegion1} and considering a fluid with isotropic pressure $p(r)$ and constant energy density $\rho$, the combination of the temporal and radial components of Einstein's equations leads to
\begin{equation}
\label{EinsteinEcuaciones}
\Phi'(r)
=
\frac{m(r)+4\pi r^3 p(r)}
{r\left[r-2m(r)\right]},
\end{equation}
where
\begin{equation}
m(r)=\frac{4\pi}{3}\rho r^3.
\end{equation}
The conservation of the energy-momentum tensor yields
\begin{equation}
\label{Conservacion}
\frac{dp(r)}{dr}
=
-\left[p(r)+\rho\right]\Phi'(r).
\end{equation}

By combining Eqs.~\eqref{EinsteinEcuaciones} and \eqref{Conservacion}, the TOV equation follows:
\begin{equation}
\label{TOVtotal}
\frac{dp(r)}{dr}
=
-\left[p(r)+\rho\right]
\frac{m(r)+4\pi r^3p(r)}
{r\left[r-2m(r)\right]}.
\end{equation}

The solution of Eq.~\eqref{TOVtotal} for constant energy density is given by
\begin{equation}
\label{PresionSchMarzo1}
p(r)
=
\rho
\frac{-1+\kappa\sqrt{\beta(r)}}
{3-\kappa\sqrt{\beta(r)}},
\end{equation}
where
\begin{equation}
\beta(r)=1-\frac{8\pi}{3}\rho r^2,
\end{equation}
and $\kappa$ is an integration constant.

The geometry is given by
\begin{align}
g_{tt}
&=
-\exp\left(2\Phi(r)\right)
=
-\kappa_{\rm sch}
\left(
1-\frac{\kappa}{3}\sqrt{\beta(r)}
\right)^2,
\label{gttSch}
\\
(g_{rr})^{-1}
&=
\exp\left(-2\Psi(r)\right)
=
\beta(r),
\end{align}
where $\kappa_{\rm sch}$ is an integration constant. In order to avoid a zero of $(g_{rr})^{-1}$ within the stellar interior, the following condition must be satisfied:
\begin{equation}
r_{\max}=R<\sqrt{\frac{3}{8\pi\rho}}.
\end{equation}

The same condition follows by requiring the argument of the square root in the solution to remain positive. Defining
\begin{equation}
M=\frac{4\pi}{3}\rho R^3,
\qquad
C=\frac{2M}{R},
\end{equation}
it follows that
\begin{equation}
\beta(r)=1-C\frac{r^2}{R^2},
\qquad
\beta(R)=1-C,
\end{equation}
and therefore $C<1$.

It is useful to mention two equivalent ways of parametrizing the solution according to the boundary condition imposed on the pressure. First, the integration constant $\kappa$ can be expressed in terms of the central pressure, $p(0)=p_C$. From Eq.~\eqref{PresionSchMarzo1}, one obtains
\begin{equation}
\label{EcuacionKappaT}
\kappa
=
\frac{3p_C+\rho}{p_C+\rho}.
\end{equation}

Consequently, Eq.~\eqref{PresionSchMarzo1} becomes
\begin{equation}
\label{presionSchPTC}
p(r)
=
\rho
\frac{
-(p_C+\rho)+(3p_C+\rho)\sqrt{\beta(r)}
}{
3(p_C+\rho)-(3p_C+\rho)\sqrt{\beta(r)}
},
\end{equation}
and
\begin{equation}
\label{gttSch1}
g_{tt}
=
-\exp\left(2\Phi(r)\right)
=
-\kappa_{\rm sch}
\left(
1-
\frac{3p_C+\rho}{3(p_C+\rho)}
\sqrt{\beta(r)}
\right)^2.
\end{equation}

Alternatively, one can impose that the pressure vanishes at the surface of the star,
\begin{equation}
p(R)=0.
\end{equation}
In this case,
\begin{equation}
\kappa=(1-C)^{-1/2},
\end{equation}
and therefore
\begin{equation}
\label{ps1}
p(r)
=
\rho
\frac{
\sqrt{1-C}-\sqrt{1-Cr^2/R^2}
}{
\sqrt{1-Cr^2/R^2}-3\sqrt{1-C}
}.
\end{equation}

Under this assumption, the central pressure is
\begin{equation}
\label{pcs}
p_C
=
\rho
\frac{\sqrt{1-C}-1}
{1-3\sqrt{1-C}}.
\end{equation}

It follows that the denominator of Eq.~\eqref{pcs} vanishes when
\begin{equation}
C=\frac{8}{9},
\qquad
\frac{M}{R}=\frac{4}{9},
\end{equation}
which corresponds to the Buchdahl limit \cite{Wald:1984rg}.

\bibliography{mybib.bib}

\end{document}